\documentclass[%
 reprint,
superscriptaddress,
 amsmath,amssymb,
 aps,
prb,
]{revtex4-2}

\usepackage{graphicx}% Include figure files
\usepackage{dcolumn}% Align table columns on decimal point
\usepackage{bm}% bold math
\usepackage{xcolor}
\usepackage[normalem]{ulem}
\usepackage{lipsum}  
\begin{document}

\preprint{APS/123-QED}

\title{Radiation Pattern Synthesis with Uniform Nonlocal Metasurfaces}% Force line breaks with \\

\author{Alexander Zhuravlev}%
 \email{a.zhuravlev@metalab.ifmo.ru}

 \author{Yury Kurenkov}%

     \affiliation{School of Physics and Engineering, ITMO University, St.~Petersburg, Russia}
     
  \author{Xuchen Wang}%
  \affiliation{International Joint Research Center for Nanophotonics and Metamaterials, Harbin Engineering University, Qingdao, China}%
  
  \author{Fedor Dushko}%
 \author{Viktor Zalipaev}%
\affiliation{School of Physics and Engineering, ITMO University, St.~Petersburg, Russia}%
\author{Stanislav Glybovski}%
%  \email{s.glybovski@metalab.ifmo.ru}
  \affiliation{School of Physics and Engineering, ITMO University, St.~Petersburg, Russia}


\date{\today}% It is always \today, today,

\begin{abstract}
One of the main applications of electromagnetic metasurfaces (MSs) is to tailor spatial field distributions. The radiation pattern of a given source can be desirably modified upon reflection on an MS having proper spatial modulation of its local macroscopic parameters. At the microscopic level, spatial modulation requires individually engineered meta-atoms at different points. In contrast, the present research demonstrates the opportunity for radiation pattern engineering in the reflection regime without using any spatial modulation. The principle consists in the deliberate tailoring of the surface impedance of an unmodulated but spatially dispersive (nonlocal) MS. A 2D synthesis problem with a magnetic line current source is solved analytically by finding a required form of the surface impedance as a function of the tangential wave vector in both visible and evanescent parts of the spatial spectrum. To prove the principle, three different pattern shapes are implemented via full-wave numerical simulations by tuning the spatial dispersion in a realistic mushroom-type high-impedance electromagnetic surface with loaded vias. This work extends the synthesis methods and the application area of spatially dispersive MSs, showing the latter as a promising platform for new types of antennas.  \end{abstract}

\maketitle

%\tableofcontents
%%%%%%%%%%%%%%%%%%%%%%%%%
\section{\label{sec:INTR}Introduction}
%%%%%%%%%%%%%%%%%%%%%%%%%

Manipulations of electromagnetic fields such as polarization control, absorption, focusing, and beam steering are conventionally implemented by means of wave plates, lenses, pyramidal absorbers, and other bulk quasi-optical devices. To date, these and even more complex functions can be achieved by using electrically thin electromagnetic metasurfaces (MSs) implemented as two-dimensional arrays with subwavelength periodicity of constituent metaatoms (reviewed, e.g. in \cite{Yu2014} and \cite{Glybovski2016}). Metasurfaces effectively behave as uniform sheets of induced surface electric and magnetic currents. Depending on the MS response, these currents are generally related to the averaged tangential electric and magnetic field components through three macroscopic surface parameters (electric impedance,
magnetic admittance, and magnetoelectric coupling coefficient) \cite{Epstein2016_arbit,Chen2020} or the polarizability tensor \cite{Radi2013}. One can distinguish penetrable and impenetrable MSs. Impenetrable (or reflective) MSs without spatial modulation are made of periodic identical meta-atoms. Such type of MSs can be used as perfect absorbers \cite{Raa'di2015}, high-impedance electromagnetic surfaces (HIS) that enable low-profile antenna reflectors \cite{Mosallaei2004,Baracco2008,Tran2010}, quasi-TEM and impedance waveguides \cite{Fei-Ran1999,Higgins2003,Luukkonen2008_Waveg}, as well as electromagnetic bandgap (EBG) structures suitable for antenna decoupling \cite{Fan2003}. Spatially modulated impenetrable MSs consist of individually engineered meta-atoms placed at different points \cite{Pozar2007}. This type of MSs can be used as metamirrors \cite{Asadchy2015} and holographic leaky-wave antennas \cite{Frong2010,Minatti2011}. 
Isotropic impenetrable MSs can be modeled using a macroscopic (averaged) boundary condition, in which only one frequency-dependent surface parameter, i.e. scalar surface impedance $Z_\text{s}$, connects tangential electric $E_\text{t}$ and magnetic $H_\text{t}$ fields at the interface as $E_{\text{t}}=Z_{\text{s}}H_{\text{t}}$. A surface impedance describing either uniform or spatially modulated MSs is most often assumed to be local. Under that assumption, the impedance boundary condition connects both tangential fields at the same point on the MS. However, in general, MSs exhibit nonlocal response, leading to the \textit{spatial dispersion} (SD) of the surface impedance, which may be weak or strong depending on the mutual coupling between adjacent and distant meta-atoms. In terms of the boundary condition, it means a certain relation between $E_{\text{t}}$ and $H_{\text{t}}$ taken at different points on the MS. In particular, SD in unmodulated impenetrable MSs under plane-wave illumination, studied hereinafter in this work, results in $Z_\text{s}$ being a function not only of frequency, but also of the incident angle.

Spatial dispersion may strongly affect the pattern of the field reflected from an impenetrable MS, which is important to consider in field pattern engineering. The first example is related to uniform artificial magnetic conductor shields. Even if the MS exhibits high surface impedance at the normal incidence of a plane wave that mimics a perfect magnetic conductor (PMC), the surface impedance can significantly alter its value with the incident angle \cite{Lukkonen2009}. As a result, the expected in-phase reflection applies only to a narrow part of the plane-wave (spatial) spectrum radiated by an antenna placed over the shield, while the rest part of the  spectrum may experience undesirable destructive interference or even couple to supported surface or leaky waves \cite{Kaipa2011}. Therefore, the MS does not behave as an ideal PMC shield for the antenna in the low-profile configuration, deteriorating the radiation pattern shape. In the second example of uniform HISs, SD is responsible for the presence of a surface-wave stopband at frequencies close to the regime of high surface impedance. Such a property is inherent in the mushroom-type HIS, but not in simple grounded dielectric slabs {\cite{Yang2003} or frequency selective surfaces (FSSs) deposited on a thin grounded dielectric slab \cite{Luukkonen2008_compar}. The third example is related to spatially modulated MSs for anomalous reflection of an incident plane wave at a given angle. In \cite{Asadchy2016} it was found that ideal anomalous reflection without polarization transformation and any auxiliary excitation waves requires SD with a specific strongly nonlocal response to the fields. All three examples show that SD is crucial to consider for precise field pattern engineering with practical (e.g. printed-circuit board (PCB)) MSs.

On the other hand, nonlocal properties of MSs can be utilized to enable new functions. Thus, in \cite{Zhirihin2017}, an MS absorber based on a mushroom-type HIS with vias loaded onto lumped circuit elements is designed to operate perfectly at two incident angles. Furthermore, structures containing loaded wire medium slab were used to build a superlens (i.e. a lens that focuses both propagating and evanescent waves) achieving a subwavelength resolution \cite{Belov2005,Kaipa2012}. Another promising function which can be implemented using nonlocal MSs is optical signal processing \cite{Kwon2018}.

Recently, it was proposed to synthesize spatially dispersive surface parameters of MSs to reach the desired operations over incident wave beams upon transmission and reflection \cite{Dugan2024}. The proposed method deriving the rational polynomial forms of the surface parameters as functions of the tangential wave vector and the corresponding higher-order boundary conditions \cite{Rahmeier2023} has been numerically confirmed for such examples as a space-plate, a spatial filter, and an all-angle perfect absorbers. 

Therefore, SD effects are significant in practical structures and may enable drastically new opportunities for field pattern synthesis. Nevertheless, systematic design principles, possible practical realizations in various frequency ranges as well as particular applications of nonlocal MSs are still to be revealed. 

In this work, we propose uniform and nonlocal MSs as functional antenna reflectors capable of shaping the radiation pattern in the desired way. With this aim, we analytically solve the boundary value problem with a magnetic line current source placed in the vicinity of an impenetrable MS with strong SD. Unlike in the previous works,
% (e.g. \cite{Tatarnikov2012,Berry1963,Arrebola2008}), 
we show the opportunity to approximate the desirable radiation pattern without any spatial modulation. Our approach is based on the deliberate tailoring of the nonlocal surface impedance of an unmodulated MS by finding and implementing its dependence on the tangential wave vector. 

The paper is organized as follows. In Sect.~\ref{subsec:BoundProb} we describe the analytical solution to the aforementioned reflection problem. Implementation of the appropriate nonlocal $Z_\text{s}$  in both visible and evanescent parts of the spatial spectrum for the practical MS design compatible with the PCB technology is discussed in Sec.~\ref{subsec:ABCP}. At the end of Sect.~\ref{sec:THR} we summarize the proposed algorithm to implement a specific shape of the radiation pattern by tailoring the parameters responsible for SD of $Z_{\text{s}}$. To prove the principle, in Sect.~\ref{sec:Results} we apply the algorithm to engineer three different radiation patterns owing to nonlocal properties of mushroom-type HIS with loaded vias and verify the results via full-wave numerical simulations. In Sect.~\ref{sec:EXP} we experimentally confirm the results for one of the three radiation patterns.
%%%%%%%%%%%%%%%%%%%%%%%%%
\section{\label{sec:THR}Theory}
%%%%%%%%%%%%%%%%%%%%%%%%%
\subsection{\label{subsec:BoundProb}Boundary value problem}

Let us consider the geometry of the two-dimensional boundary value problem shown in Fig.~\ref{Fig1:BP}. 
%%%%%%%%%
\begin{figure}
  \centering
  \includegraphics[width=0.9\columnwidth]{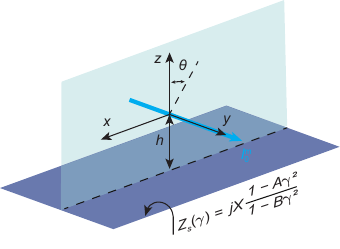} 
  \caption{Sketch of the considered two-dimensional boundary value problem. Infinite in $y$ direction magnetic line source is placed at a height $h$ above an unmodulated MS. The MS is characterized by impenetrable spatially dispersive surface impedance $Z_\text{s}(\gamma)$, where $\gamma=k_x/k$ is the tangential component $k_x$ of the wave vector normalized to the wavenumber $k$ of free space.}
  \label{Fig1:BP}
\end{figure}
%%%%%%%%%
The infinite line of magnetic current flowing along $y$ axis is located above the uniform, isotropic, and spatially dispersive MS with zero thickness, which occupies the plane $z=-h$. The field has transverse-magnetic (TM) polarization, which means that only the $H_y$, $E_x$, and $E_z$ field components are nonzero. In this work we assume the time dependence in the form $e^{j\omega t}$. The incident field vectors created by the source can be written as \cite[par.~5.4]{Felsen1994}:

%%%%%%%%%
\begin{equation}
\begin{array}{cc}
 \textbf{H}^{\text{\text{inc}}}=-\frac{I^{\text{m}}_0k}{4\eta}H^{(2)}_0(kr)\textbf{y}_0,& \textbf{E}^{\text{\text{inc}}}=\frac{j\eta}{k}\nabla\times\textbf{H}^{\text{\text{inc}}}
\end{array}
\label{eq:Fields}
\end{equation}
%%%%%%%%%
where $I^{\text{m}}_0$ is the magnetic current amplitude, $H^{(2)}_0(kr)$ is Hankel function of the second kind and zeroth order, $k$ is the free space wavenumber, $\textbf{y}_0$ is the unit vector along $y$ axis,  $\eta$ is the impedance of free space, and $r=\sqrt{x^2+z^2}$.

We solve the problem in the spatial spectral domain and represent the incident and reflected fields as their continuous plane-wave spectra. The transmitted field is assumed to be zero. With respect to each incident spatial harmonic, having normalized tangential component of the wave vector $\gamma=k_x/k$, the response of the spatially dispersive MS is modeled with impenetrable surface impedance $Z_\text{s}(\omega,\gamma)$, where $\omega$ is an arbitrarily chosen frequency of interest. The values $\gamma=\sin(\theta) < 1$, with $\theta$ being the incident and reflection angle, correspond to propagating plane waves, while the values $\gamma \geq 1$ correspond to evanescent waves. To simplify the problem and maintain generality, the approximation of $Z_\text{s}(\omega,\gamma)$ is applied as described below. Following \cite{Bankov2022} and \cite{Dugan2024,Rahmeier2023} we represent $Z_\text{s}(\omega,\gamma)$ as a rational function of $\gamma$. The latter can be obtained by applying Pade approximation, which is physically motivated by modeling the frequency response of the MS with a set of Lorentzian resonances \cite{Maci2005} dependent on the wave vector, as discussed in \cite{Rahmeier2023}. This approximation can capture possible poles and zeros of $Z_\text{s}(\omega,\gamma)$ with respect to $\gamma$. Assuming a practical realization of the MS as a uniform periodic structure of subwavelengh meta-atoms, we note that the possible values of $\gamma$ of excited waves are limited by periodicity $p$. Therefore, our approximation should be correct for $0 \le |\gamma| < \gamma_{\text{max}}$, where $\gamma_{\text{max}} \approx \pi/kp$ (i.e., should be valid for the visible or evanescent part of the spectrum). Assuming the MS to be lossless and reciprocal, and that $Z_\text{s}(\gamma)$ at a single frequency exhibits no more than one zero and no more than one pole in the approximation range, it can be simplified by keeping terms in the rational function up to the second order as follows:
%%%%
\begin{equation}
    Z_\text{s}(\gamma)=jX\frac{1-\sum_{n=1}^{\infty} A_n\gamma^{2n}}{1-\sum_{m=1}^{\infty} B_m\gamma^{2m}}\approx jX\frac{1-A\gamma^2}{1-B\gamma^2}
    \label{eq:aprox}
\end{equation}
%%%
where $X$ is the surface reactance value at normal incidence, $A$ and $B$ are the coefficients that define the angular position of the zero and pole, respectively. Note that the first and higher odd powers of $\gamma$ are omitted in (\ref{eq:aprox}) assuming reciprocal properties of the MS \cite{Bankov2022}. For a lossless structure, all coefficients are frequency-dependent real numbers. It should be noted that setting $B$ to zero reduces \eqref{eq:aprox} to the particular case of second-order averaged boundary conditions in the same form as derived for the penetrative grid impedance of thin wire meshes by M.~I.~Kontorovich et al. \cite{Kontorovich1962} (see also in \cite[Ch.4]{TretyakovAMAE}).

The $y$ component of the magnetic field from \eqref{eq:Fields} may be represented as an integral decomposition to plane waves via Fourier transform, as, e.g. in \cite{Tatarnikov2012,Tretyakov20023}. The corresponding spectral function for the incident field reads:
%%%%%%%%%%%%%
\begin{equation}
   \hat{H}^{\text{inc}}=-\frac{I^{\text{m}}_0}{4\eta}\frac{2}{\sqrt{1-\gamma^2}}e^{-jk\sqrt{1-\gamma^2}|z|}
\end{equation}
%%%%%%%%%%%%%
where $(\hat{...})$ denotes the field in the spatial spectrum domain. To find the field reflected from the MS, we impose the impenetrable boundary condition \eqref{eq:aprox} at the interface $z=-h$ for each wave of the spectrum, assuming that the field in the region $z<-h$ is zero. We obtain:
%%%%%%%%%%%%%
\begin{equation}
   (\hat{E}^{\text{inc}}+\hat{E}^{\text{ref}})\sqrt{1-\gamma^2}=-Z_\text{s}(\gamma)(\hat{H}^{\text{inc}}+\hat{H}^{\text{ref}})\Bigr\rvert_{z = -h}
   \label{eq:bound_cond_for_E}
\end{equation}
%%%%%%%%%%%%%
where $\hat{E}^{\text{inc}}$ and $\hat{E}^{\text{ref}}$ are the complex amplitudes of incident and reflected fields in the spatial spectrum domain. Eliminating the electric fields by substituting $\hat{E}^{\text{inc}}=\eta \hat{H}^{\text{inc}}$ and $\hat{E}^{\text{ref}}=-\eta \hat{H}^{\text{ref}}$ into \eqref{eq:bound_cond_for_E}, one can derive the reflection coefficient for the magnetic field:
%%%%%%%%%%%%%
\begin{equation}
    \rho^{H} = \frac{\hat{H}^{\text{ref}}}{\hat{H}^{\text{inc}}}=-\rho=\frac{\eta^{\text{TM}}-Z_\text{s}(\gamma)}{\eta^{\text{TM}}+Z_\text{s}(\gamma)}
    \label{eq:reflcoef}
\end{equation}
%%%%%%%%%%%%%
where $\eta^{\text{TM}}=\eta\sqrt{1-\gamma^2}$ is the characteristic impedance of the TM wave which is the ratio between the electric and magnetic field components of an incident plane wave tangential to the plane $z=-h$ and $\rho=\hat{E}^{\text{ref}}/\hat{E}^{\text{inc}}$ is the reflection coefficient for the electric field. Substituting the approximation \eqref{eq:aprox} in \eqref{eq:reflcoef} one obtains:
%%%%%%%%%%%%%
\begin{equation}
    \rho=\frac{jX(1-A\gamma^2)-\eta\sqrt{1-\gamma^2}(1-B\gamma^2)}{jX(1-A\gamma^2)+\eta \sqrt{1-\gamma^2}(1-B\gamma^2)}
    \label{eq:refcoef_X_A_B}
\end{equation}
%%%%%%%%%%%%%

Possible zeros of the denominator with respect to $\gamma$ of the expression correspond to propagation factors of supported surface waves (if the root is a real number) or leaky waves (if the root is a complex number) \cite{Silveirinha2008}. Since in our study we aim to engineer the radiation pattern, excitation of surface waves is considered as an adverse effect because they can influence the radiation pattern in the case of a finite shield due to edge diffraction \cite{Roudot1985}. Equating the denominator of \eqref{eq:refcoef_X_A_B} to zero and making substitution $\sqrt{1-\gamma^2}=\xi$, one can obtain a cubic equation:
%%%%%%%%%%%%%
\begin{equation}
B\eta \xi^3+jXA\xi^2+j\eta \xi (1-B)+jX(1-A)=0
    \label{eq:disp}
\end{equation}
%%%%%%%%%%%%%
To avoid surface wave propagation, the imaginary part of the roots $\gamma_{\text{sw},i}$ of that equation (with $i$ being the root number) should be positive (assuming $e^{-jk\xi z}$ dependence of the reflected field on the coordinate $z$), because this condition leads to the nonphysical waves, which are impossible to excite in real structures. The spectral function of the reflected magnetic field in the upper half-space ($z>-h$) may be written as follows:
%%%%%%%%%%%%%
\begin{equation}
    \hat{H}^{\text{ref}}=-\rho \hat{H}^{\text{inc}}\Bigr\rvert_{z = -h}e^{-jk\sqrt{1-\gamma^2}(z+h)}
\end{equation}
%%%%%%%%%%%%%
The corresponding distributions of the incident and reflected magnetic fields (namely their $y$ components) in coordinate space can be found by inverse Fourier transform: 
%%%%%%%%%%%%%
\begin{equation}
        H^{\text{inc/ref}}=\frac{k}{2\pi}\int_{-\infty}^{+\infty}\hat{H}^{\text{inc/ref}}e^{jk \gamma x}d\gamma
\label{eq:invF}
\end{equation}
%%%%%%%%%%%%%
Once we are interested in the radiation pattern (i.e. the far-field dependence on the observation angle $\theta$), the inverse Fourier transform can be asymptotically calculated by the method of steepest descent \cite[Ch. 4]{Felsen1994}:
%%%%%%%%%%%%%
\small
\begin{eqnarray}
    H^{\text{inc}}_{r\to\infty}(\theta)&\approx& \sqrt{2\pi}e^{j\frac{3\pi}{4}}\frac{kI^{\text{m}}_0}{4\pi\eta}\frac{e^{-jkr}}{\sqrt{kr}}
    \label{eq:fields_xz}  \nonumber \\
    H^{\text{ref}}_{r\to\infty}(\theta)&\approx&-\rho(\theta)\sqrt{2\pi}e^{j\frac{3\pi}{4}}\frac{kI^{\text{m}}_0}{4\pi\eta}\frac{e^{-jkr_1}}{\sqrt{kr_1}} + \\ 
    &+&2\pi j \sum_{i}{\text{Res}(\hat{H}^{\text{ref}},\gamma_{sw,i})} \nonumber 
\end{eqnarray}
\normalsize
%%%%%%%%%%%%%
where $r_1=\sqrt{x^2+(z+2h)^2}$ is the position of the imaginary source located in the mirror position below the MS, and the sum of the residues corresponds to the poles of the expression in \eqref{eq:refcoef_X_A_B}. It is possible to further simplify \eqref{eq:fields_xz}, as the observation point is in the far-field zone, by applying the following approximation: $e^{-jkr_1}/\sqrt{kr_1}\approx e^{-jk(r+2h\cos(\theta))}/\sqrt{kr}$. The total field equals the sum of $H^{\text{inc}}_{r\to\infty}(\theta)$ and $H^{\text{ref}}_{r\to\infty}(\theta)$:
%%%%%%%%%%%%%
\begin{equation}
    H^{\text{tot}}_{r\to\infty}(\theta)=\sqrt{2\pi}e^{j\frac{3\pi}{4}}\frac{kI^{\text{m}}_0}{4\pi\eta}\frac{e^{-jkr}}{\sqrt{kr}}\left(1-\rho(\theta)e^{-jk2h\cos(\theta)}\right)
\label{eq:rad_pat}
\end{equation}
%%%%%%%%%%%%%
where the sum of the residues can be omitted provided that the aforementioned condition to the roots of \eqref{eq:disp} is satisfied, and in the absence of excited surface waves the far-field asymptotics results in an expanding cylindrical wave. The radiation pattern being the normalized angle-dependent multiplier of that wave can be engineered by adjusting the angular dependence of the reflection coefficient, which is, in turn, governed by the spatial dispersion of the surface impedance. For the class of impenetrable nonlocal MSs modeled using the second-order boundary conditions \eqref{eq:aprox}, the radiation engineering implies adjusting the values of coefficients $X$, $A$, and $B$. The total field distribution is modified thanks to the changing phase of the reflection coefficient with respect to particular spatial harmonics. It is worth noting that in the given class of MSs one can employ only three degrees of freedom with a constraint of the condition to the roots of \eqref{eq:disp}.
%%%%%%%%%%%%%
%%%%%%%%%%%%%
\subsection{\label{subsec:ABCP}Implementation of spatially dispersive boundary condition}
%%%%%%%%%%%%%
%%%%%%%%%%%%%
Next, we discuss one of the possible ways to implement the nonlocal boundary condition \eqref{eq:aprox} with particular values of the coefficients $X$, $A$, and $B$ in the isotropic impenetrable structure with strong SD, which is compatible with the PCB technology. 

One of the most popular artificial impedance structures is the mushroom-type HIS, composed of small square patches connecting to a ground plane with vertical metal vias, proposed in \cite{Sievenpiper1999} with an equivalent-circuit approach describing the frequency dispersion of its local surface impedance neglecting SD. More recent and accurate analytical models of the HIS composed of metallic patches placed on a grounded dielectric slab \cite{Luukkonen2008} and of the mushroom-type HIS \cite{Lukkonen2009} include SD. In both models, the analytical description of wave propagation within the substrate of the patches allows us to correctly describe the frequency and angle dependence of $Z_{\text{s}}$. Considering TM wave excitation, the impedance of the mushroom-type HIS shows almost no angular dependence as a result of the stop-band effect in the medium of metal vias, while the structure of patches without vias shows strong variation with an incident angle. To control SD in the mushroom-type HIS, the authors of \cite{Kaipa2011} suggest adding loads (i.e., lumped circuit elements with impedance value $Z_{\text{L}}$)  connecting the bottom ends of the vias to the ground plane, as shown in Fig.~\ref{Fig:MHIS}. A normally incident wave does not interact with the vias, and, consequently, the loads do not influence the value of $Z_\text{s}(\gamma)$ for $\gamma=0$. However, under oblique incidence, the currents flow through the loads. Therefore, while $X$ can be adjusted by changing the period $p$ and gap with $g$ between patches, as well as the thickness $l$ and relative dielectric permittivity $\epsilon_h$ of the substrate, the coefficients $A$ and $B$ responsible for SD strongly depend on $Z_{\text{L}}$.
%%%%%%%%%%%%%
\begin{figure}
  \centering
  \includegraphics[width=0.9\columnwidth]{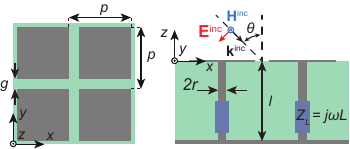} 
  \caption{Mushroom-type HIS with loaded vias: top view (left) and cross section (right).}
  \label{Fig:MHIS}
\end{figure}
%%%%%%%%%%%%%

% --------
%  In conjunction with \eqref{eq:rad_par_final} that expressions are suitable for engineering the MS. In other words, the microstructure parameters of meta-atoms providing a radiation pattern with desirable shape (in frames of the available degrees of freedom) can be easily found analytically.
% ----

Despite the fact that there is an analytical solution for $Z_{\text{s}}(\gamma)$ of the considered structure derived in \cite{Kaipa2011}, its expression is rather complicated and does not take the form of a rational function. As proposed in \cite{Rahmeier2023,Dugan2024}, the function of the MS response (the surface impedance in our case) should be approximated with a rational function containing polynomials of the desirable order. Here, we derive the second-order approximation by retaining terms up to $\gamma^2$ in the Taylor expansions of the numerator and denominator of $Z_{\text{s}}(\gamma)$ at $\gamma=0$.  Taylor expansion at the point $\gamma=0$ was chosen because a far-field distribution depends mostly on the behavior of $Z_{\text{s}}(\gamma)$ in the range  $\gamma<1$ (the range of visible incident angles). This approximation provides us closed-form expressions for $X$, $A$, and $B$ as functions of the microstructure parameters explained in Fig.~\ref{Fig:MHIS}. Appendix~\ref{appendix_A} summarizes the expressions and discusses the method for their derivation. Together with \eqref{eq:rad_pat}, these expressions are suitable for engineering the MS. In other words, the microstructure parameters of meta-atoms providing a radiation pattern with desirable shape (in frames of the available degrees of freedom) can be easily found analytically. Fortunately, from \eqref{eq:app_aprox_end} and \eqref{eq:app_sub_aprox} it is seen that $X$ does not depend on $Z_{\text{L}}$, the value of $A$ does not depend on $g$, while $B$ depends on all parameters of the structure. One may conclude that disadvantageously $A$ and $B$ cannot be independently controlled for the given structure. However, as we show in the next section, it is possible to find particular combinations of the coefficients to achieve several practically important radiation pattern shapes.

In view of the above discussion, we suggest the following algorithm for engineering the radiation pattern:
%%%%%%%%%%%%%
\begin{enumerate}
    \item select the radiation pattern shape to be realized;
     \item find the combination of $X$, $A$, and $B$, which minimizes the mean squared error (or other criteria) between the target radiation pattern and that calculated through \eqref{eq:rad_pat} imposing a constraint by ensuring that roots of the dispersion relation \eqref{eq:disp} are non-physical; 
    \item select which parameters of the microstructure should be adjusted for the considered MS type;
    \item calculate a data set of $X$, $A$, and $B$ coefficients using \eqref{eq:app_aprox_end} and \eqref{eq:app_sub_aprox} by varying the chosen structure parameters;
    \item find the triplet of $X$, $A$, and $B$ in the data set which is the closest to the combination from step 2.
\end{enumerate}
%%%%%%%%%%%%%

The meta-atom parameters determined using the above algorithm provide a good first-order solution to the MS synthesis problem. In fact, the presence of parasitics in the meta-atom (e.g. the reactance of the connection between the lumped element and the bottom end of the metal via) results in a necessary correction of the chosen parameters. As we do in the next section, this correction can be performed numerically by converging from the above first-order solution and checking the numerically calculated dependence $Z_{\text{s}}(\gamma)$ (at least for incident angles of plane waves with $0 \le \gamma < 1$). Alternatively, the influence of the parasitics could be taken into account semi-analytically as proposed in \cite{Kaipa2011} by introducing an effective load impedance.

\section{\label{sec:Results}Radiation pattern engineering} 
%%%%%%%%%%%%%
Here we present the results of engineering three different radiation patterns in the system consisting of a magnetic line source placed above the mushroom-type HIS with loaded vias. In each case, the algorithm discussed in Sect.~\ref{sec:THR} is used to realize one target radiation pattern shape applicable in wireless communication and positioning systems as follows: 
%%%%%%%%
\begin{enumerate}
    \item with constant level of radiation in the range of angles $\theta$ from $-60^\circ$ to $60^\circ$ with a sharp cut-off outside the range (\textit{$\Pi$-shaped});
    \item with a symmetrical smooth dip around the normal direction with the far-field level proportional to $1/\cos(\theta)$ in the range of angles $\theta$ from $-40^\circ$ to $40^\circ$  (\textit{Secant});
    \item with nulls created at specific angles $\theta =\pm20^\circ$ (\textit{Nulls}).
\end{enumerate}

To be consistent with PCB technology, for three corresponding realizations of the loaded mushroom-type HIS, we use a commercial Rogers RO4003C substrate with relative permittivity of $\epsilon_h=3.38(1-j0.0027)$ and thickness $l=1.524$~mm. Also, for all three MS samples we fix the periodicity ($p=0.13\lambda_0$), where $\lambda_0$ is the wavelength at the frequency of interest (10 GHz) and the diameter of the vias ($2r=0.4$~mm). To be able to model the MS with a homogeneous surface impedance, the distance between the source and the MS should be kept greater than $(1\ldots2)p$. We take $h=1.5p\approx 0.2\lambda_0$. To influence the value of $X$, we change $g$, and the last degree of freedom to simultaneously modify $A$ and $B$ is the value of loads. During analytical calculations, we found that to achieve all three goals, the loads should be inductive, i.e. $Z_{\text{L}}=j\omega L$.

% Figure \ref{Fig:XAB} shows the dependence of $X$, $A$ and $B$ coefficients on the value of $L$ and width of the gap between patches $g$ calculated using the expressions from Appendix~\ref{appendix_A}. Colormaps in Fig~\ref{Fig:XAB} are calculated for $0 \le L \le 1$ nH, and $0.02 \le g/p \le 0.42$. According to Fig.~\ref{Fig:XAB}(a) $X$ changes the sign around $g/p\approx0.07$ which corresponds to the resonance of the patch array on the grounded dielectric slab at the normal incidence. Coefficients $A$ and $B$ change the signs around $L\approx0.24$~nH (see Fig.~\ref{Fig:XAB}(a) and (b)), but coefficient $B$ also changes the sign around $g/p\approx0.07$. This property allows $A$ and $B$ to  be chosen such that these coefficients have either same or opposite signs. The results from Fig.~\ref{Fig:XAB} are used to find $X$, $A$, and $B$ coefficients which provide the desired radiation pattern shapes. The optimal values of $X$, $A$, and $B$ in this work were found using the mean squared error criteria to approximate the desired pattern shapes and are depicted in the Fig.~\ref{Fig:XAB} with colored circles.  The red and orange circles correspond to $\Pi$-shaped and Secant radiation patterns consequently, while the blue circles correspond to Nulls radiation pattern. By finding abscissa and ordinate of each point, it is possible to obtain the geometrical parameters for each MS implantation. The geometric parameters of the MSs are summarized in Table~\ref{tab:geom_param_0}.

Fig.~\ref{Fig:XAB} shows the dependence of the coefficients $X$, $A$, and $B$ on the value of $L$ and the ratio $g/p$ calculated using the expressions of Appendix~\ref{appendix_A}. The color maps in Fig.~\ref{Fig:XAB} are calculated for $0 \le L \le 1$~nH, and $0.02 \le g/p \le 0.42$. The results of Fig.~\ref{Fig:XAB} are used to find coefficients $X$, $A$, and $B$ that provide the desired shapes of the radiation pattern. The optimal values of $X$, $A$ and $B$ in this work are found using the mean squared error criteria to approximate the desired pattern shapes, and their determined values are displayed in Fig.~\ref{Fig:XAB} with colored circles. The red and orange circles correspond to the $\Pi$-shaped and Secant radiation patterns consequently, while the blue circles correspond to the Nulls radiation pattern. By finding the abscissa and ordinate of each circle's position, it is possible to obtain initial (first-order approximated) geometrical parameters for each implantation of the loaded mushroom-type HIS. The coefficients $X$, $A$, and $B$, as well as the geometric parameters of meta-atoms corresponding to the considered shapes of the radiation pattern are summarized in Table~\ref{tab:geom_param_0}.
%%%%%%%%%%%%%
\begin{figure*}
  \centering
  \includegraphics[width=0.9\textwidth]{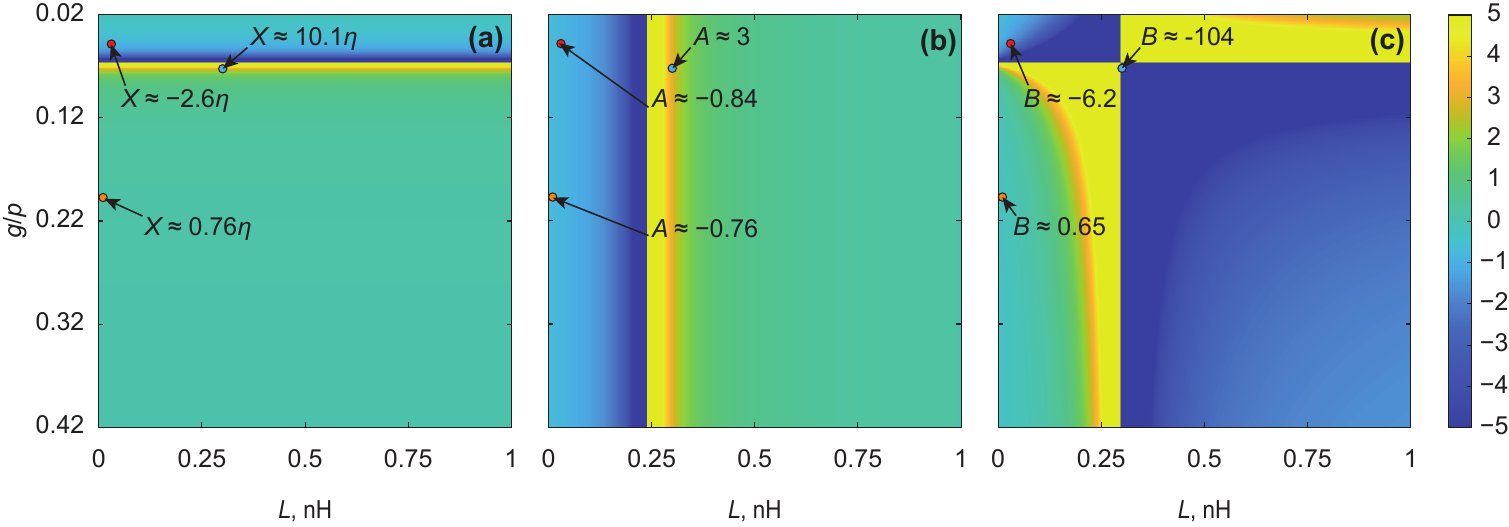} 
  \caption{The coefficients $X/\eta$ (a), $A$ (b), and $B$ (c) vs. the load inductance and the ratio of $g/p$. Colored circles depict the three solutions with the desired shapes of the radiation pattern: the red and orange circles correspond to the $\Pi$-shaped and Secant patterns consequently, while the blue circles correspond to the Nulls pattern. }
  \label{Fig:XAB}
\end{figure*}
%%%%%%%%%%%%
\begin{table}
\caption{The analytically predicted coefficients $X$, $A$, $B$, and the corresponding geometric parameters of idealized meta-atoms (with lumped loads) found using the expressions of Appendix~\ref{appendix_A} for three different shapes of the radiation pattern.}
\label{tab:geom_param_0}
\begin{ruledtabular}
\begin{tabular}{lcccccr}
Pattern & $X/\eta$ & $A$ & $B$ & $g/p$& $L$, nH  \\
\colrule
$\Pi$-shaped & $-2.6$ & $-0.84$ & $-6.2$ & $0.05$ & $0.03$ & \\
Secant & $0.76$ & $-0.76$ & $0.65$ & $0.20$ & $0.01$ & \\
Nulls & $10.1$ & $3.0$ & $-104$ & $0.072$ & $0.30$ &
\end{tabular}
\end{ruledtabular}
\end{table}
%%%%%%%%

After finding values of the coefficients and the analytically predicted geometric parameters, a realistic (fabrication-ready) structure is to be built using numerical software. An important issue is to implement the inductive lumped loads within the capabilities of the PCB technology. Another issue is related to parasitics that exist in the realistic structure. As stated in \cite{Kaipa2011}, it can be addressed semi-analytically by extracting the values of the effective load impedance from additional numerical simulations. Fig.~\ref{Fig:Sim_Exp}(a) shows the top perspective view of the practical geometry of the unit cell bounded by four periodic symmetry planes and two Floquet ports modeled in CST Studio Suite (CST). Inductive loads are implemented as structural inductance based on three types of narrow copper strips that connect the bottom end of a via to the ground plane. The strips are etched in the layer of the ground plane and have different shapes in three considered cases (see  Fig.~\ref{Fig:Sim_Exp}(b)). In order to obtain the geometric parameters of each structural inductance corresponding to the required limped-load inductance (see the right column of Table~\ref{tab:geom_param_0}), numerically calculated lookup tables can be used, which are illustrated by the graphs in Fig.~\ref{Fig:table}. Note that for each structural inductance type the primary geometric parameter ($w$ or $\varphi$) to be swept is selected, while the other parameters ($s$, $b$ and $d$) are kept constant.
\begin{figure}
\centering
  \includegraphics[width=0.85\columnwidth]{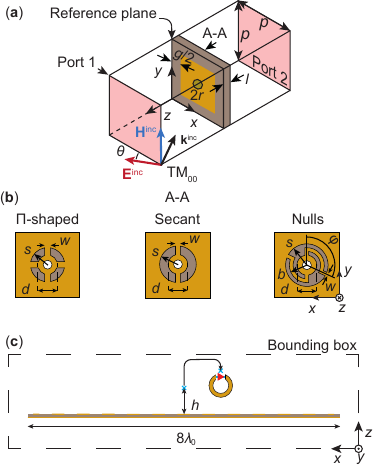} 
  \caption{Simulation setup for numerical calculation of $Z_{\text{s}}(\gamma)$ in CST (top perspective view of one periodic unit cell) (a). The patterns etched in the copper ground plane to implement inductive loads (b). Simulation setup for a MS excited by a magnetic line current source for numerical calculation of a radiation pattern in $xz$-plane with CST (c).}
  \label{Fig:Sim_Exp}
\end{figure}
\begin{figure}
  \centering
  \includegraphics[width=0.9\columnwidth]{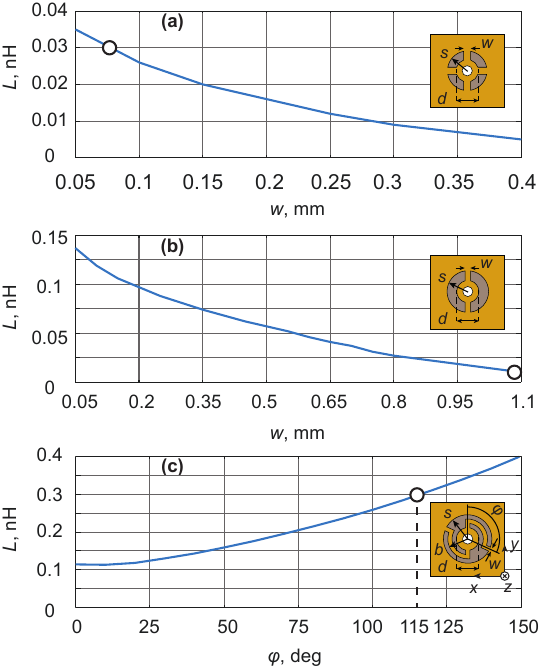} 
  \caption{Correspondence between the lumped inductance of a load and the primary geometric parameters to sweep of structural inductance (lookup tables) for the: (a) $\Pi$-shaped, (b) Secant, and (c) Nulls radiation pattern. Markers denote the values of the initial meta-atom geometry (before fine tuning).}
  \label{Fig:table}
\end{figure}

Next, to compensate for the effect of parasitics (existing even for idealized lumped loads), one should further perform a fine parametric tuning of $g$ and ($w$ or $\varphi$ depending on the shape of the structural inductance). In this step, the goal is to make the numerically calculated curve $Z_{\text{s}}(\gamma)$ as close as possible to that predicted by \eqref{eq:aprox} for the $X$, $A$ and $B$ listed in Table~\ref{tab:geom_param_0}. It should be noted that to reach the goal the surface impedance is important to calculate
either for propagating or evanescent waves (for $\gamma$ up to $\pi/kp$) using a suitable order of the incident Floquet mode. Analyzing the dependencies of Fig.~\ref{Fig:XAB} may help simplify the parametric tuning. The value $Z_{\text{s}}(0)$ could be refined by adjusting $g/p$, then the rest of the curve could be refined by adjusting the parameters of the structural inductance. 

For our three target radiation patterns, the analytically predicted curves $Z_{\text{s}}(\gamma)$ according to Fig.~\ref{Fig:XAB} are compared with those calculated numerically for the practical geometry with structural load inductance in Fig.~\ref{Fig:IMP}. For this comparison we show the initial numerical curves (before fine tuning) corresponding to the geometric meta-atom parameters estimated using the expressions of Appendix~\ref{appendix_A} and lookup tables of Fig.~\ref{Fig:table}. In addition, in Fig.~\ref{Fig:IMP} we show the final numerical curves, corresponding to the fine-tuned  parameters of the practical meta-atoms. The latter are listed in Table~\ref{tab:geom_param}.
\begin{figure}
  \centering
  \includegraphics[width=0.9\columnwidth]{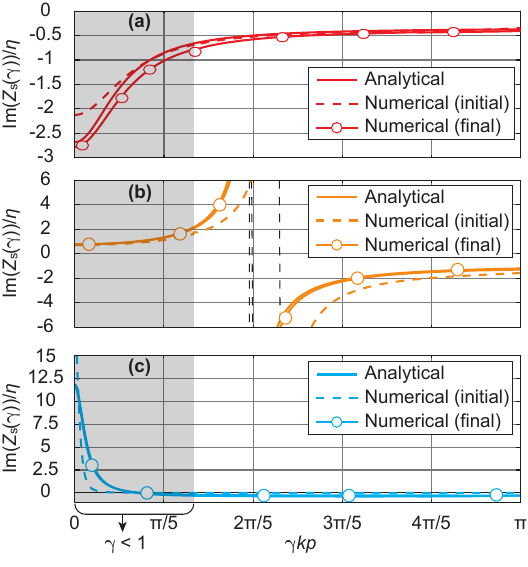} 
  \caption{Dependencies of $Z_{\text{s}}$ on normalized tangential component of the incident field wave vector for the desired MS. Plot (a) for $\Pi$-shaped radiation pattern, plot (b) for Secant radiation pattern, and plot (c) for Nulls radiation pattern. Legend captions: \textit{"Analytical"} denotes curves calculated analytically using formula \eqref{eq:aprox} with coefficients of Table~\ref{tab:geom_param_0}; \textit{"Numerical (initial)"} denotes initial curves calculated numerically for the practical geometry with structural load inductance (before fine tuning) corresponding to the geometric meta-atom parameters estimated using the expressions of Appendix~\ref{appendix_A} and calculated lookup tables; \textit{"Numerical (final)"} denotes final curves calculated numerically for the practical geometry with structural load inductance (after fine tuning).
}
  \label{Fig:IMP}
\end{figure}
\begin{table}
\caption{Geometric parameters of practical meta-atoms with structural inductance after fine tuning for three different shapes of the radiation pattern.}
\label{tab:geom_param}
\begin{ruledtabular}
\begin{tabular}{lcccccr}
Pattern & $g/p$ & $d$, mm & $s$, mm & $w$, mm & $b$, mm & $\phi, ^\circ$ \\
\colrule
$\Pi$-shaped & $0.053$ & $0.80$ & $0.70$ & $0.25$ & - & - \\
Secant & $0.195$ & $0.80$ & $0.75$ & $0.20$ & - & - \\
Nulls & $0.080$ & $0.80$ & $0.85$ & $0.15$ & $0.70$ & $115$
\end{tabular}
\end{ruledtabular}
\end{table}
%%%%%%%%
It can be seen that the final numerical curves agree well with
the predictions of \eqref{Fig:XAB}, especially
in the range important for radiation pattern engineering ($\gamma<1$ highlighted with gray fill), while the initial curves provide a good first-order estimation.
%Also, to illustrate the precision of the second-order approximation of the SD used for the boundary conditions, the same figure also contains the curves $Z_{\text{s}}(\gamma)$ calculated using equation (5) of \cite{Kaipa2011}. For $\gamma<1$ the approximation is good enough for all cases, while there is a considerable error in the spectral position of the pole for the Secant radiation pattern case.

%HERE
Using the final parameters in Table~\ref{tab:geom_param} the near-field and far-field patterns of the linear magnetic current source in the presence of a finite-size nonlocal MS (in its fabrication-ready form) can be calculated via full-wave numerical simulations.
Fig.~\ref{Fig:Sim_Exp}(c) depicts the corresponding simulation model used with three considered meta-atom parameters.
For the two-dimensional problem we model only one linear array of meta-atoms with a length of $8\lambda_0$, where $\lambda_0$ is the free-space wavelength. The structure is placed in the bounding box between two zero-phase-delay periodic boundary conditions along the $y$ axis. 
The parameter $h$ of the figure is the distance between the top plane of the MS (the plane of patches) and the phase center of the magnetic line source implemented as a metal tube uniformly prolonged along the $y$ axis. The tube with a deep subwavelength cross section has a lateral radiating slot with a uniform lumped port installed in it, as shown in the inset of Fig.~\ref{Fig:Sim_Exp}(c). Like in the analytical model, we choose $h=0.2\lambda_0$ and show all the field patterns at 10 GHz. 

Figure \ref{Fig:results} shows the comparison between the results of analytical calculations and full-wave simulations: the distributions of total magnetic field amplitudes (first row) and radiation patterns (second row). The analytical field distributions in Fig.~\ref{Fig:results} were obtained by numerically solving the inverse Fourier transform \eqref{eq:invF} for $H^{\text{inc}}+H^{\text{ref}}$, while the analytical radiation patterns $F(\theta)$ shown with red curves were obtained by normalizing the absolute value of \eqref{eq:rad_pat} as a function of $\theta$ to its maximum. The blue curves show the numerically calculated radiation patterns in CST, while the green curves show numerical results obtained by solving the boundary
value problem in Comsol. The black dashed lines depict the target radiation patterns. Despite simulating finite MSs, one can see that full-wave results are in good agreement with the analytical predictions for all three target pattern shapes, which validates our approach. The numerically calculated radiation patterns indeed reproduce the main features of the target shapes, that is, the flatness in the angular range from $-60^{\circ}$ to $60^{\circ}$ is achieved in Fig.\ref{Fig:results}~(b), the secant profile is achieved in the range from $-40^{\circ}$ to $40^{\circ}$ in panel (d), while the target position of the nulls ($\pm 20^{\circ}$) is achieved in panel (f). Moreover, from the analytical and numerical near-field distributions, which agree well, one can notice that there is no field localization near the $z=0$ plane, which confirms the absence of propagating surface waves.
 %%%%%%%%
\begin{figure*}
  \centering
  \includegraphics[width=0.95\textwidth]{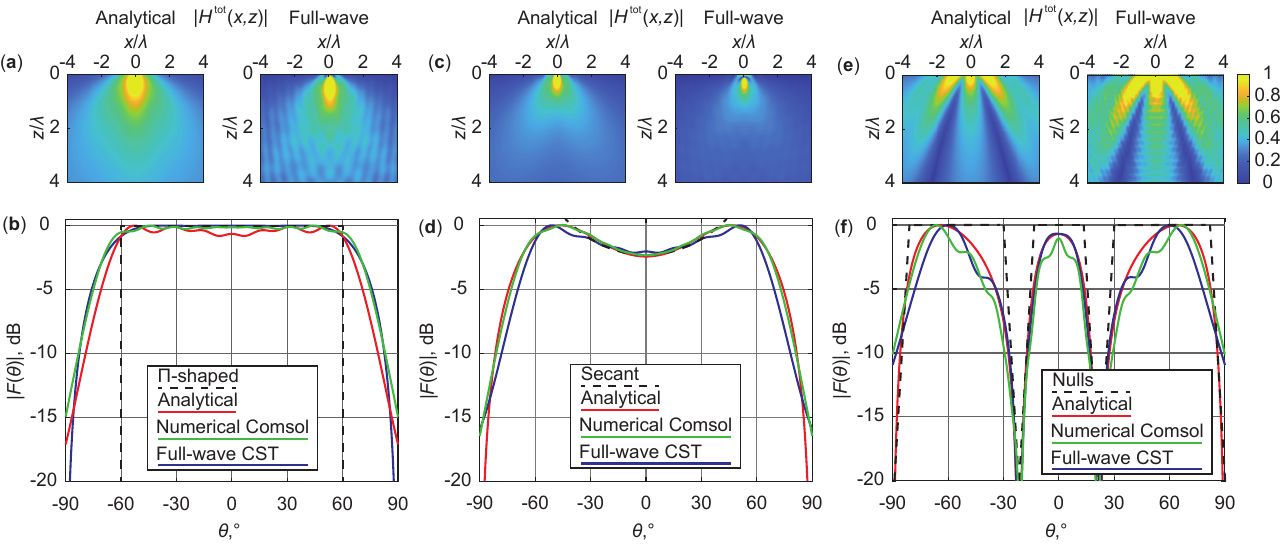} 
  \caption{The results of full-wave simulations obtained for finite-size and fabrication-ready MSs compared to the analytical predictions for the corresponding infinite nonlocal boundary model for the $\Pi$-shaped (a,b), Secant (c,d), and Nulls (e,f) target shapes of the radiation pattern. The near-field distributions are shown in the first row, while radiation patters are shown in the second one. Target radiation patterns are displayed with black dashed curves. All data are normalized by maximum.}
  \label{Fig:results}
\end{figure*}
%%%%%%%%

\section{\label{sec:EXP}Experimental validation}

In order to additionally validate the proposed radiation pattern synthesis approach, an experiment was made with one of the three MSs designed above, i.e. with one providing the Secant radiation pattern. The sketch and photograph of the experimental sample are shown in Fig.~\ref{Fig:experiment}(a) along with the top and bottom views of $3\times 3$ meta-atoms manufactured with the parameters given in the second row of Table~\ref{tab:geom_param}. In the E-plane ($xz$-plane, in which the radiation pattern was measured), the size of the MS was the same as in full-wave simulations ($8\lambda_0=24$~cm). The entire length of the manufactured PCB was larger (28 cm) to create two fixing flanges. The size of the MS and the entire sample in the H-plane ($yz$-plane) was arbitrary set at 19 cm.

The sample was fabricated using a dual-side
RO4003C substrate with thickness $l = 1.524$ mm and relative permittivity $\epsilon = 3.38(1-j 0.0025)$. The magnetic line source was implemented as a flat open-ended horn based on a thin parallel plate waveguide. The horn is placed in the central $yz$ plane (orthogonal to the MS) and fed with a shielded transition to a coaxial connector located below the MS. The parallel plate waveguide was made of a 0.25-mm-thick FR-4 substrate with dual-sided metallization. The horn was inserted into a thin slot in the middle of the MS PCB to obtain the desired height $h$ of its open end above the plane of the patches, as shown in Fig.~\ref{Fig:experiment}(a). On the bottom side, both metal sheets of the waveguide were soldered to the ground plane of the MS.

The setup for far-field measurements in
an anechoic chamber with the sample installed on top of a foam rotating table is shown in Fig.~\ref{Fig:experiment}(b). To measure the radiation pattern, the SMA connector of the source horn representing the magnetic line current was connected to the first port of Rohde\&Schwarz ZVB20 vector network analyzer (VNA), while the second port was connected to a TEM horn probe placed 4 m away. The manufactured sample was rotated in the E-plane
and the measured coefficient $S_{21}$ was stored for each rotation angle $\theta$ from $-90^{\circ}$ to $90^{\circ}$ with a step of $1^{\circ}$. In Fig.~\ref{Fig:experiment}(c) a good agreement between the results of measurements and full-wave simulations can be observed. 
\begin{figure*}
  \centering
  \includegraphics[width=0.95\textwidth]{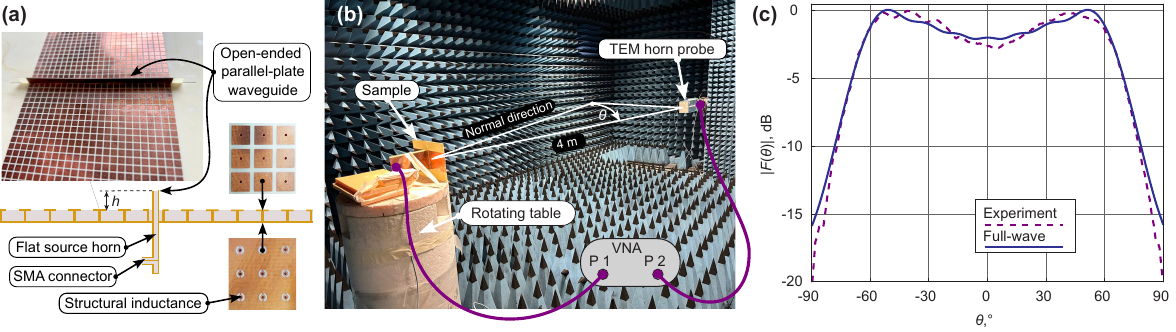} 
  \caption{The experimental verification of the possibility of the radiation patter engineering: (a) sketch and photographs of the manufactured sample; (b) experimental setup for radiation pattern measurements; (c) measured radiation in comparison with the full-wave calculated one at 10 GHz for the Secant radiation pattern shape.}
  \label{Fig:experiment}
\end{figure*}
%%

%%%%%%%%
\section{Conclusion}
%%%%%%%%
In this paper, we proposed the use of nonlocal impenetrable MSs as artificial antenna reflectors that assist in approximating the desired shape of the radiation pattern. We showed that on-demand modification of the radiation pattern is possible with a new principle, i.e. by engineering strong SD of a uniform MS. Properly tailored SD allows different spatial spectrum harmonics (both propagating and evanescent waves) created by the given source to reflect with different predefined phases without dissipation, thereby building the desired total field distribution. In contrast to previous approaches, we employ no spatial modulation, which simplifies the synthesis and fabrication procedures as all meta-atoms are identical. Another new advantage of the proposed reflectors is that the source can be freely moved parallel to the MS plane without changing the radiation pattern. Indeed, unlike conventional spatially modulated MS reflectors designed for a specific source position, here only the height of the source is important.

To demonstrate the proposed approach, we developed a semi-analytical synthesis algorithm that results in a fabrication-ready MS geometry compatible with the PCB technology. The novelty consists in the possibility of following the entire route from the desired shape of the radiation pattern to the particular microstructure of meta-atoms through the analysis of the required SD. 
The algorithm employs the second-order approximation in nonlocal impedance boundary conditions as well as previously derived homogenization theory of a loaded mushroom-type HIS simplified to relatively simple closed-form expressions. Moreover, a new analytical formula was derived that allows controlling the absence of propagating surface waves to avoid parasitic edge diffraction effects for finite-size reflectors. This contribution fills the gap between previously known analytical modeling methods of nonlocal impedance boundaries and useful potential applications of SD. 

The proposed approach was validated on three examples with different target shapes of the radiation pattern, all excited by a linear magnetic current source. The target patterns considered are useful in antennas for wireless communication and positioning systems. In all cases, the built analytical model allows one to predict the law of spatial dispersion for the surface impedance and to obtain a good estimation of the microstructure parameters of PCB-compatible meta-atoms. The influence of parasitic reactances and the inaccuracy of the second-order approximation could be compensated by means of fine tuning of the microstructure parameters. The resulting numerical and experimental field patterns obtained for finite-size MS samples were found to be in good agreement with the analytical predictions for an unbounded nonlocal impedance sheet. In turn, the possibility of avoiding propagating surface waves was proven by comparing the analytical and full-wave field distributions near to the MS plane.

Note that the target shapes of the radiation pattern could not be perfectly fitted, especially at glancing angles for the $\Pi$ and Secant shapes, as well as for angles between the nulls for the Nulls shape. The
key limitation of the proposed algorithm is that there are very few degrees of freedom to design an arbitrary radiation pattern. In fact, in the second-order boundary condition used, there are only three real coefficients: $X$, $A$, and $B$. Moreover, despite the fact that in the considered loaded mushroom-type HIS there are enough microstructure parameters to realize the three considered target shapes of the radiation pattern, independent adjustment of $X$, $A$, and $B$ is still impossible in this structure. As a result, the variety of achievable $Z_{\text{s}}(\gamma)$ functions is also limited.

The results of this work can be further developed to introduce new types of low-profile antennas with unique radiation properties. Thus, the possibilities of engineering higher-order nonlocal boundary conditions and their realization in practical MS structures for more precise engineering of near- and/or far-field patterns are a part of future studies.

\begin{acknowledgments}
The authors thank D.V. Tatarnikov for useful discussions.\\
\end{acknowledgments}

%%%%%%%%
%%%%%%%%
\appendix
%%%%%%%%
%%%%%%%%

\section{\label{appendix_A} APPROXIMATION COEFFICIENTS FOR LOADED MUSHROOM-TYPE HIS}

In this appendix, closed-form expressions for the coefficients $X$, $A$, and $B$ of the approximate second-order boundary condition \eqref{eq:aprox} are derived for the mushroom-type HIS with loaded vias from an analytical solution for $Z_{\text{s}}(\gamma)$ obtained in \cite{Kaipa2011}.

The coefficients can be derived using the formula for the TM-polarized plane wave reflection coefficient from the considered structure (Eq.~(5) in \cite{Kaipa2011}), which is valid for an arbitrary tangential component $k_x=\gamma k$ of the wave vector. With this aim, the following steps are to be taken. First, Eq.~(5) from \cite{Kaipa2011} for the reflection coefficient should be expressed in a form of \eqref{eq:reflcoef} from this work, and the expression for $Z_\text{s}(\gamma)$ should be obtained in a fractional form.
Next, one should apply Taylor expansion separately to the numerator and denominator of $Z_\text{s}(\gamma)$ with respect to $\gamma$ at $\gamma=0$ keeping terms up to $\gamma^2$.
Finally, the approximation of $Z_{\text{s}}(\gamma)$ should be reshaped to the form of the right-hand side of \eqref{eq:aprox} to obtain the expressions for the coefficients $X$, $A$, and $B$.

Following the above algorithm, the result can be written as: 
\begin{eqnarray}
\label{eq:app_aprox_end}
&X&(p,\epsilon_h,l,g)=-\frac{\eta\alpha_0}{\eta X_g\alpha_0+k\beta_0} \nonumber\\
&A&(p,\epsilon_h,l,r,Z_L)=\frac{\alpha_1}{\alpha_0}\\
&B&(p,\epsilon_h,l,g,r,Z_L)=\frac{\eta X_g\alpha_1+k\beta_1}{\eta X_g\alpha_0+k\beta_0} \nonumber 
\end{eqnarray}
%%%%%
The meaning of $\alpha_0$, $\alpha_1$, $\beta_0$, $\beta_1$ in \eqref{eq:app_aprox_end} is explained as follows: 
\begin{widetext}
\begin{eqnarray}
\label{eq:app_sub_aprox}
    \alpha_0&=&\frac{k_\text{TEM}k_p^2}{2\xi k^2}\left(-2\xi+\xi l^2 \tau^2+2l\tau^2\right)\sin(k_\text{TEM}l) \nonumber\\
    \alpha_1&=&-\frac{k_\text{TEM}l}{2\xi}\left(k_p^2(\xi l+4)-2k_\text{TEM}^2\right)\sin(k_\text{TEM}l)-l\tau^2\cos(k_\text{TEM}l)  \\ 
    \beta_0&=&\left[\epsilon_h\frac{k_p^2}{k^2}\left(-\frac{l^2}{2}\tau^2+1\right)-\epsilon_h\frac{k_p^2 l}{\xi k^2}\tau^2\right]\cos(k_\text{TEM}l) \nonumber\\
    \beta_1&=&2\epsilon_h + \left[\frac{\epsilon_h}{2}(k_\text{TEM}^2 l^2 - 2) + \frac{\epsilon_h k_p^2 l}{\xi}\right]\cos(k_\text{TEM}l) + \left[\frac{\epsilon_h k_\text{TEM}}{2\xi}(\tau^2 l^2 - 2) + \frac{\epsilon_h l}{k_\text{TEM}}(k_p^2 - 2k_\text{TEM}^2)\right]\sin(k_\text{TEM}l)
    \nonumber 
\end{eqnarray}
with
\begin{eqnarray}
        \xi&=&j\omega C Z_{L} \nonumber \\
    \tau&=&\sqrt{k_\text{TEM}^2 - k_p^2} \nonumber 
\end{eqnarray}
\end{widetext}
where $k_{\text{TEM}}=k\sqrt{\epsilon_h}$ is the propagation constant of transverse
electromagnetic (TEM) wave in the wire medium, modeling the dielectric substrate with vertical vias, $\epsilon_h$ is the relative dielectric permittivity of the substrate (host medium for the vias), $k_p=\sqrt{(2\pi/p)^2/\log[p^2/4r(p-r)]}$ is the plasma wavenumber of the wire medium; $X_g=(\epsilon_h + 1)(kp/\eta \pi)\log[\csc(\pi g/2 p)]$ is the grid susceptance of the patch array; $C=2\pi \epsilon_0 \epsilon_h/\log[(p^2/4r(p-r)]$ is capacitance per unit length of the wire medium \cite{Maslovski2009}; $Z_L$ is the impedance of the via load, and $\omega$ is the angular frequency.

% \begin{table}[h]
% \caption{Final geometric parameters of the MSs for each radiation pattern implementation.}
% \label{tab:geom_param}
% \begin{ruledtabular}
% \begin{tabular}{lcccccr}
% Pattern & $g/p$ & $d$, mm & $s$, mm & $w$, mm & $b$, mm & $\phi, ^\circ$ \\
% \colrule
% $\Pi$-shaped & $0.053$ & $0.8$ & $0.70$ & $0.25$ & - & - \\
% Secant & $0.195$ & $0.8$ & $0.75$ & $0.20$ & - & - \\
% Nulls & $0.080$ & $0.8$ & $0.85$ & $0.15$ & $0.7$ & $115$
% \end{tabular}
% \end{ruledtabular}
% \end{table}
% %%%%%%%%

%%%%%%%%%%%%
% \begin{figure*}[hb!]
%   \centering
%   \includegraphics[width=0.95\textwidth]{Figs/Fig.table_ind.pdf} 
%   \caption{}
%   \label{Fig:XAB}
% \end{figure*}


% \nocite{*}
\bibliography{apssamp}
\end{document}